\documentclass[11pt,tightenlines,
floats,aps,amsmath,amssymb,nofootinbib,prd]{revtex4}

\usepackage{setspace}
\usepackage{amsmath,amssymb,amsfonts,amsthm}
\usepackage{graphicx}
\usepackage{enumerate} % advanced enumerate environment
\usepackage{colordvi} % for color text

\newcommand{\be}{\nopagebreak[3]\begin{equation}}
\newcommand{\ee}{\end{equation}}
\newcommand{\ba}{\nopagebreak[3]\begin{eqnarray}}
\newcommand{\ea}{\end{eqnarray}}
\newcommand{\nn}{\nonumber \\}

\newcommand{\f}{\frac}

\def\tr{{\rm Tr}}

\begin{document}
\title{ \Large 
%{\tt DRAFT}\\[5mm]
 Stepping out of Homogeneity in Loop Quantum Cosmology}

\author{Carlo Rovelli{\it ${}^a$}, Francesca Vidotto{\it ${}^{ab}$} \vspace{0.2cm}}

\affiliation{\small\it ${}^a$Centre de Physique Th\'eorique de Luminy\footnote{Unit\'e mixte de recherche (UMR 6207) du CNRS et des Universit\'es de Provence (Aix-Marseille I), de la M\'editerran\'ee (Aix-Marseille II) et du Sud (Toulon-Var); laboratoire affili\'e \`a la FRUMAM (FR 2291).}, Case 907, F-13288 Marseille, EU\\
\small\it ${}^b$Dipartimento di Fisica ``Galileo Galilei'', Universit\`a di Padova, via F.\,Marzolo 8, I-35131 Padova, EU \vspace{0.2cm}}

\date{\small May 29, 2008}

\begin{abstract}

\noindent
We explore the extension of quantum cosmology outside the homogeneous approximation,  
using the formalism of loop quantum gravity.   We introduce a model where {\em some} of
the inhomogeneous degrees of freedom are present, providing a tool for describing general fluctuations of quantum geometry near the initial singularity. We show that the dynamical structure of the model reduces to that of loop quantum cosmology in the Born-Oppenheimer approximation.  This result corroborates the assumptions that ground loop cosmology, sheds light on the physical and mathematical relation between loop cosmology and full loop quantum gravity, and on the nature of the cosmological approximation.  Finally, we show that  the non-graph-changing Hamiltonian constraint considered in the context of algebraic quantum gravity  provides a viable effective dynamics within this approximation.
\end{abstract}

\maketitle

\section{Introduction}

Loop quantum cosmology (LQC) provides the most successful physical application of loop gravity, and one of the most promising avenues towards the possibility of an empirical test of quantum gravity \cite{ashtekar, bojowald}. (For a recent review of quantum cosmology, see \cite{Kiefer:2008sw}.) In particular, the possibility of a fully consistent quantum description of ``big-bang physics'', and the robustness of the bounce prediction, represent a clear advance in our understanding of quantum gravitational physics within this theoretical framework.  This impressive success opens a number of physical and mathematical questions:    
(i) Can we include inhomogeneities? Inhomogeneities and their quantum fluctuations play a fundamental role in the currently fashionable cosmological scenario.  Inhomogeneities can in principle be re-inserted at a later cosmological epoch, restricting the analysis of the Planck epoch to the sole homogeneous degrees of freedom---but is this approximation viable? After all, what is very interesting is precisely to understand the configuration of the full fluctuating quantum geometry near the singularity itself, about which very little is known (see \cite{fotini}). In other words:  
(ii) Can we describe the actual quantum state of the geometry near the initial singularity, beyond the homogeneous approximation? Perhaps this state could even teach us something {\em directly} about the emergence of the physical inhomogeneities of our universe.   
(iii) What is the true relation between full loop quantum gravity \cite{lqg,lqg2} and LQC?  The question has been addressed repeatedly \cite{thiemann} and concerns have been raised on whether the two theories are truly consistent---some simple minded ways of interpreting their relation have even been recently rigorously proven incorrect \cite{fleishhach}.

We address here all these questions. We do so by first analyzing the nature of the approximation on which cosmology itself --classical or quantum-- is based.  This is neither a low-energy nor a high-energy approximation, since cosmology appears to describe well very large distance features of our universe as well as its behavior in much higher energy-density regimes.  The analysis leads us to the idea that the full theory may be consistently expanded by adding degrees of freedom one by one, starting from the cosmological ones.  Accordingly, we define an approximated dynamics of the universe, inhomogeneous but truncated at a finite number of degrees of freedom, and we discuss its regime of validity.  (For previous works on inhomogeneities in LQC, see
\cite{MartinBenito:2008ej,  Dittrich:2007jx, Bojoino}.  See also \cite{Bojoino2}.)
This approximation includes and extends conventional cosmology, without however including the full infinite dimensional field theory. We work in the compact case, which is conceptually simpler ---that is, we assume that the topology of the spacial universe is that of a three-sphere (a possibility which is still compatible with observations, and, according to some \cite{Ellis}, it is even favored by them).  The approximation we take can be intuitively interpreted as a truncation of all degrees of freedom to a finite order in a multipolar expansion of the fields on the topological three-sphere.  These degrees of freedom can be described using a fixed 3d compact triangulation $\Delta_n$, formed by $n$ tetrahedra. 

The quantum kinematics of this system turns out to be described by the truncation of loop quantum gravity obtained by restricting the spin-network states to those based on a graph equal (or contained in) to the graph defined by the dual of $\Delta_n$. Within the approximation considered,  the quantum dynamics can be described by the non-graph-changing version of the Hamiltonian constraint \cite{thiemanncit} that has been recently considered in the context of algebraic quantum gravity \cite{algebraic}.  Thus, non-graph-changing Hamiltonian constraint plays here the role of an {\em effective} dynamics, as originally suggested by Thiemann\footnote{``Maybe one could call the operator as formulated in this section an effective operator". Ref. \cite{thiemanncit}, Sect 5.2.}. In this way, one can try to define a quantum cosmological model for any given triangulation $\Delta_n$, with a number of degrees of freedom that increases with the complexity of $\Delta_n$. 

We study here in detail the simplest nontrivial case, based on a triangulation $\Delta_2$ formed by the minimal triangulation of a 3-sphere: two tetrahedra glued along each face. The corresponding dual graph is formed by two nodes joined by four links.  We show that the model is well defined and in particular the constraint algebra closes.  We write the state space, the quantum operators and the Hamiltonian constraints of this model explicitly. The model represents a non  homogeneous quantum universe, where, say, we do not consider just the overall spacial average of a scalar field, but also its dipole moment.  This provides a well-defined ``first step out of homogeneity" in loop quantum cosmology.   We call this model a ``dipole cosmology", but the name should not be taken literally, as the gravitational degrees of freedom that are described are more ``quadrupolar" that dipolar: the spatial universe is split into two hemispheres (represented by the two tetrahedra), separated by a closed surface $\Sigma$; but $\Sigma$ splits in turn into {\em four} large surfaces (the four triangles bounding the tetrahedra), whose areas provide degrees of freedom that can be roughly thought as capturing the geometry of a 3d ellipsoid.

We then ask in which sense LQC  is contained in the larger model.  We argue that a proper way of addressing the problem is  to interpret LQC as a first-order Born-Oppenheimer approximation. This is the approximation in which the effect of the inhomogeneities on the dynamics of the scale factor is small, as the effect of the electrons on the dynamics of the nuclei is small, in the Born-Oppenheimer approach to molecules \cite{BO}. (For other utilizations of the Born-Oppenheimer approximation in this context see \cite{BOP}.)  We show concretely that taking the order-zero Born-Oppenheimer approximation of the  $\Delta_2$ model yields precisely the structure of the LQC dynamics. In particular, we recover the characteristic structure of the LQC Wheeler-DeWitt equation, defined as a 3-terms finite-difference equation in the scale factor.  This derivation provides a prototype for understanding how LQG is contained in full loop quantum gravity. In particular, we derive here the quantization of the LQC $\mu$ parameter directly from LQG, without need of an explicit recourse to the area gap argument \cite{ashtekar}.

The models defined here, and in particular the  $\Delta_2$ dipole cosmology, are finite-dimensional quantum theories, that can be used to describe the inhomogeneous quantum geometry near the initial singularity and its quantum fluctuations, to any given arbitrary order.  We leave the analysis of these models to further investigations.

The paper is organized as follows. In Section 2, we discuss the nature of the approximation we take and its viability.    In Section 3, we define the quantum models for arbitrary $\Delta_n$. The $\Delta_2$ model is described in some detail in Section 4. The Born-Oppenheimer approximation is introduced in Section 5, where we show how the structure of LQC can be recovered.  In Section 6  we conclude pointing out some implications of the results presented, in particular  their relation with Regge calculus cosmology \cite{barrett}, and with the approximation used in computing background independent $n$-point functions \cite{propagator}.

\section{Approximations in cosmology}

Modern cosmology was born with Einstein's 1917 seminal paper \cite{einstein}, where Einstein states the {\em cosmological principle}, according to which the dynamics of a homogeneous and isotropic space approximates the one of our real universe.   A certain vagueness lingers around the precise status of this principle, presented under a variety of different lights in the literature. The question of whether the universe approaches homogeneity at large scale is an empirical question: the later evolution of observational cosmology appears to be corroborating it, as well as finding some precise quantitative limits to large scale homogeneity.  Einstein's principle is therefore a hypothesis, in the healthy tradition of hypothetical-deductive science.  But what precisely is the hypothesis, and why is it relevant?  The universe is obviously \emph{not} homogeneous at every scale and not \emph{exactly} homogeneous at \emph{any} scale, except at its very largest scale, where it is homogeneous by definition.   Why can we neglect the effect of the inhonomogeneities on the dynamics of this largest scale, in a nonlinear theory like general relativity, where {\em all}  scales are coupled?  In fact, \emph{this} is precisely the hypothesis put forward by Einstein in 1917: that the universe happen to be in a state where the effect of the inhomogeneities on the dynamics of its largest scale, described by the scale factor, can be neglected in a first approximation.  More generally, that the universe is in a configuration where the effect on the
longest-wavelength of the interaction with the shorter-wavelengths is negligible.   The significance of this hypothesis is becoming particularly clear today, since a number of contrary hypotheses are being explored, such as, in particular, the very intriguing possibility that the measured cosmological constant could be --in full or in part-- the result of the effect of these shorter wavelengths on the largest scale (see for instance \cite{B}). In other words, the cosmological principle is the hypothesis that a certain approximation scheme is viable in general relativity, and that the universe happens to be in the regime where this approximation scheme is effective. 

What precisely is this approximation? One is tempted to say that it is simply a long-distance one: it is defined by cutting-off the modes with wavelength shorter than a certain size $L$.  But this is imprecise, because $L$ varies with the size itself of the scale factor, and can be also very small. The situation is easier to analyze in the context of a spatially closed universe; let us therefore assume  in the following that we are in this context.  Let $a^3(t)$ be the volume of the universe at the cosmological time $t$.  Then the approximation on which cosmology is based is to neglect the dynamics of the wavelengths $\lambda$ shorter than $a(t)$.  But $a(t)$ itself can be large as well as small.  If $n\sim\frac{a}\lambda$, the cosmological approximation is the first order term, $n=1$, in an expansion in small $n$, which does not necessarily mean large $\lambda$. 

Consider the next terms in the expansion, say $n=2, 3, ...$ and so on.  An immediate consequence of the cosmological principle (or a natural extension of the same) is the hypothesis that the dynamics of the long wavelengths modes of the universe is only weakly affected by its highest modes.  We can implement this expansion by approximating the geometry of the universe not just by a maximally symmetric space, as in standard cosmology, but rather by a geometry described by a finite number of degrees of freedom. These can be labeled by the elements of a triangulation $\Delta_n$ formed by a finite number $n$ of tetrahedra.   
If a maximally symmetric space represents the gravitational degrees of freedom of the universe averaged over the largest possible scale, the degrees of freedom of a fixed triangulation can be interpreted as a representation of the gravitational degrees of freedom of the universe, averaged over the large scales, up to a certain degree in a mode expansion.  In the following, we consider the classical and quantum description of the dynamics of a model of the universe defined in this manner. That is, we construct an \textit{effective} theory where wave lengths $\lambda <\lambda_0\sim  \frac{a}{n}$ are neglected. 

\section{The model}

\subsection{Classical theory}

Fix an (oriented) triangulation $\Delta_n$ of a (topological) three-sphere, formed by $n$ tetrahedra $t$ glued by their triangles. We label the triangles with an index  $f$ (``$f$" for \emph{face}) that runs from 1 to $2n$ (the number if faces is twice the number of tetrahedra).  Associate a group element $U_f\in SU(2)$ and a $su(2)$ algebra element $E_f$ to each oriented triangle $f$. We take the convention that to the face $f^{-1}$ obtained inverting the orientation of $f$ is associated the group element 
\be\label{Umeno1}
U_{f^{-1}}=U^{-1}_{f}
\ee
and the algebra element 
\be\label{Emeno1}
E_{f^{-1}}= - \ U_{f}^{-1}E_{f}U_{f}.
\ee
If $\tau_i,\  i=1,2,3$ is a basis in $su(2)$, we write $E_f=E_f^i\tau_i$. 
We take $U_f$ and $E_f$ as phase space variables of a dynamical system, with the conventional Poisson brackets structure of a canonical lattice $SU(2)$ Yang-Mills theory, that is
\ba 
\{  U_f,U_{f'} \} &=& 0,  \label{uno}\\
\{  E_f^i,U_{f'} \} &=&  \delta_{\!f\!f'} \ \tau^i U_{f},  \label{due}\\ 
\{  E_f^i,E_{f'}^j \} &=& - \  \delta_{\!f\!f'}\  \epsilon^{ijk} E_f^k.  \label{tre}      % aggiunto segno meno
\ea
In other words, the phase space is the cotangent bundle of $SU(2)^{2n}$ with its natural symplectic structure.  Let the dynamics of the system be defined by two sets of constraints: the gauge constraints
\be\label{gauge}
  G_t \equiv  \sum_{f\in t} E_f \sim 0,
\ee
where the sum is over the four faces of the tetrahedron; and the Hamiltonian constraints       
% aggiunta nota
%\footnote{The Hamiltonian constraint written in such a manner gives gives the good limit once the gauge constraint has been taken. In fact at the first order the holonomy is  $U\sim \exp\int_{\alpha}A\sim 1\!\! 1 + |\alpha|^2 F_{ab} +O(|\alpha|^4 A^2)$. If we insert it in our expression in \ref{clham}, we obtain $V_t C_t \equiv \sum_{ff'\in t} \tr[(E_{f'}E_{f}] + \sum_{ff'\in t} \tr[|\alpha|^2 F_{ab}E_{f'}E_{f}] \sim 0$, where the latter term has the right continuum limit and the former is zero once one uses the gauge constraint: $\sum_{ff'\in t} \tr[E_{f'}E_{f}]= \tr[(\sum_{f\in t}E\_{f})(\sum_{f'\in t}E_{f'})]= 0$.}
\be\label{clham}
C_t \equiv  V^{-1}_t \sum_{ff'\in t} \tr[U_{\!f\!f'}E_{f'}E_{f},
% -U_{\!f'\!f}E_fE_{f'}
] \sim 0
\ee
where the sum is over the couples of distinct faces at each tetrahedron, 
$U_{\!f\!f'}=U_{f}U_{f_1}U_{f_2} ... U_{f'}^{-1}$
where \mbox{ $l_{\!f\!f'}=\{f,f_1,f_2,...,f'^{-1}\}$} is the \emph{link}
%va bene  il link cos" scritto per uniformitˆ con le altre definizioni nuove?
of the oriented faces around the edge where $f$ and $f'$ join, and  
\be\label{volume}
V_t^2=    \frac14 \sum_{ff'f'\!'\in t} \tr[E_fE_{f'}E_{f'\!'}].
\ee
where the sum is over the four unordered triplets of distinct faces at the tetrahedron and $\{f,f',f''\}$ has positive orientation. Notice that  $V_t^2 =  \tr[E_fE_{f'}E_{f'\!'}]$ because of (\ref{gauge}). This concludes the definition of the dynamical systems we want to consider.  In Section IV we study one of these  systems in more detail, and we make sure the constraint algebra closes.  We leave the analysis of the constraint algebra in the general case for future developments. 

This dynamical system can be interpreted as a cosmological approximation to the dynamics of the geometry of a closed universe. To see this, consider real Ashtekar fields  $A^i_a(x)$ and $E^{ia}(x)$, with their standard Poisson algebra (see for instance \cite{lqg}), on a 3d surface  $\Sigma$ with the $S_3$ topology. (The index $a$ is a 3d (abstract) tangent index.)  Let $\Delta_n$ be a triangulation of  $\Sigma$ and $\Delta^*_n$ a dual of the triangulation. We interpret $U_f$ as the parallel transport of the Ashtekar connection $A_a(x)$ along the link $e_f$ of $\Delta^*_n$ dual to the triangle $f$, and $E_f$ as the flux $\Phi_f$ of the Ashtekar's electric field $E^a(x)$ across the triangle $f$, {\em parallel transported  to the center of the tetrahedron}.    That is, $E_f=U_{f_1}\Phi_fU^{-1}_{f_1}$ and  $E_{f^{-1}}=-U^{-1}_{f_2}\Phi_fU_{f_2}$ where $U_f=U_{f_1}U_{f_2}$ and  $U_{f_1}$ and $U_{f_1}$ are the holonomies of the two segments $e_{f_1}, e_{f_2}$ into which the face $f$ cuts the link $e_f$. Then the Poisson brackets of $U_f$ and $E_f$ defined in these manner turn out to be precisely (\ref{uno},\ref{due},\ref{tre}).   In particular, notice that the origin of (\ref{tre}) is the fact that $\Phi_f$ is parallel transported to the center of each tetrahedron. (Notice also that (\ref{tre}) follows from (\ref{uno}), (\ref{due}) and the Jacobi identity.)

The gauge constraint (\ref{gauge}) generates  the correct internal gauge transformations on these variables.%
\footnote{In particular, writing $G[\lambda]:=2\sum_t\tr[\lambda_t G_t]$ where $\lambda_t\in su(2)$, the infinitesimal gauge transformation of $U_f$ is $\delta U_f=\{U_f,G[\lambda] \}=\lambda_{t_1}U_f-U_f\lambda_{t_2}$ where $(t_1, t_2)$ are the two tetrahedra separated by $f$.}   If the triangulation is sufficiently fine, (\ref{clham}) approximate the Ashtekar's (euclidean part of the) Hamiltonian constraint $\tr[F_{ab}E^aE^b]/\sqrt{det E}\sim 0$, where $F_{ab}$ is the curvature of $A_a$. 
Notice the absence of the second term of the usual discretization  $\tr[(U_{\!f\!f'}-U_{\!f\!f'}^{-1})E_{f'}E_{f}]$ of the Hamiltonian constraint.  The $U_{\!f\!f'}^{-1}$ is usually subtracted in order to subtract the first term in the small-curvature expansion  $U\sim \exp\int_{\alpha}A\sim 1\!\! 1 + |\alpha|^2 F_{ab} +O(|\alpha|^4 A^2)$; but the subtraction is not needed because this term does not contribute to $C_t$ thanks to (\ref{gauge}). We do not know if this observation has already been made in the literature. An explicit calculation shows that $C_t$ is real. 

Finally, $V_t$ is (proportional to) the volume of the tetrahedron $t$ and we call $V=\sum_t V_t$ the total volume of space.  We take for convenience $8/3\pi G_{Newton}$, the speed of light and $\hbar$ to be unit, and we choose the Immirzi parameter $\gamma=1$ for simplicity; what follows needs to be extended to the more interesting case of real $\gamma$ and full Hamiltonian constraint.

The constraint (\ref{clham}) corresponds to the non-graph-changing version of the Hamiltonian constraint, as the one utilized in algebraic quantum gravity \cite{algebraic}.  This is the only viable alternative in the present context, where we have reduced the degrees of freedom of the gravitational field to a fixed number. This version of the constraint approximates the classical Hamiltonian constraint if $U_{\!f\!f'}$, namely  the parallel transport of $A_a$ along the loop $\alpha$ dual to the link $l_{\!f\!f'}$, approximates  $1\!\! 1  + |\alpha|^2 F_{ab}$. It is important to notice that this happens not only if the length of the loop is small, but also for large loops if $A_a$ is small. 
%(because $U\sim \exp\int_{\alpha}A\sim 1\!\! 1 + |\alpha|^2 F_{ab} +O(|\alpha|^4 A^2)$)
Hence near flat spacetime the approximation can be good even for coarse triangulations.  Misunderstanding of this fact has generated the erroneous idea that low-curvature spacetime needs to be approximated by fine triangulations. 

Alternatively, one can interpret the data $(\Delta_n, U_f, E_f)$  as a description a piecewise flat Regge geometry, where the curvature of the connection is concentrated on the edges of $\Delta$.  The quantity $U_{\!f\!f'}$ gives then the curvature at the corresponding edge.  
However, observe this interpretation is slightly misleading here, since the variables of the model are better understood as macroscopic quantities averaging over local degrees of freedom. Therefore the flatness of the individual tetrahedra does not need to be take literally.

It's simple to couple a family of multifingered ``clock" variables, one per node.
The simplest choice \cite{ashtekar} is an (ultra-local) scalar field
\footnote{Ultralocal scalar fields are distinguished by the independent temporal development of the field at each spacial point.}
 with a value $\phi_t$ and conjugate momentum $p_{\phi_t}$ at each node, with the overall Hamiltonian constraint given by 
\be\label{ham3}
 \frac{1}{V_t} \sum_{ff'\in t} \tr[U_{\!f\!f'}E_{f'}E_{f}
 %-U_{\!f'\!f}E_fE_{f'}
 ]  + \frac{\kappa}{2V_t}\  p_{\phi_t}^2  \sim 0,  
\ee
where $\kappa$, proportional to the Newton constant $G$, determines the matter-gravity coupling.   The role of this field is double.  First, it keeps track of evolution in a background-independent manner, namely it models a physical clock. 
Second, it represents in a simplified manner the matter content of the universe.  Replacing this field with a more realistic description is viable here: ultralocality can be eliminated adding a difference term; while Yang-Mills and fermion fields have a particularly straightforward description in this language \cite{lqg}.

\subsection{Quantum theory}

The quantization of the model is immediate. Following what is done in lattice QCD, a quantum representation of the observable algebra (\ref{uno}-\ref{tre}) is provided by the Hilbert space $H_{aux}=L_2[SU(2)^{2n}, dU_f]$ where $dU_f$ is the Haar measure.  The states have the form $\psi(U_f)$.  The operators $U_f$ are diagonal and the operators $E_f$ are the left invariant vector fields on each $SU(2)$. The operators $E_{f^{-1}}$ turn then out to be the right invariant vector fields.  The operator associated to the volume $V_t$ turns out to be the standard loop-quantum-gravity volume operator that is  constructed in terms of $E_f$. The states that solve the gauge constraint (\ref{gauge}) are labeled by SU(2) spin networks on the graph $\Delta_n^*$, which has a node for each tetrahedron and a link for each face of $\Delta_n$.    A basis of these is given by states $|j_f, \iota_t\rangle$, where $f=1,...,2n$ and $t=1,...,n$ range over the links and the nodes of the graph. These are defined by
\be
 \psi_{j_{\!f}\iota_t}\!(U_f) \equiv\langle U_f | j_f, \iota_t \rangle \equiv  \otimes_f\  \Pi^{(j_f)}(U_f)  \cdot  \otimes_t \ \iota_t  
\ee
where $\Pi^{(j)}(U)$ are the matrix elements of the spin-$j$ representation of SU(2) and ``$\cdot$" indicates the contraction of the indices of these matrices with the indices of the intertwiners $\iota_t$ dictated by the graph $\Delta_n^*$. For details, see \cite{lqg}.

With a scalar field, the Hilbert space becomes $H_{aux}=L_2[SU(2)^{2n}, dU_f]\otimes L_2[R^n]$, with a (generalized) basis $ | j_f, \iota_t,\phi_t \rangle $ and the states can be written in the form
\be
\psi(j_f, \iota_t,\phi_t ) \equiv \langle  j_f, \iota_t, \phi_t  | \psi \rangle. 
\ee
In this basis the operator $\phi_t$ is diagonal while $p_{\phi_t}=-i\frac{\partial}{\partial \phi_t}$.

If all constraints are first class, they can be quantized  \emph{\`a la}  Dirac. The quantum Hamiltonian constraint  can be defined in two alternative forms. The first, \emph{\`a la} Thiemann, is obtained rewriting (\ref{clham}) in the Thiemann's form 
\be\label{hamth}
C_t  =  \  \sum_{ff'f'\!'\in t} \epsilon^{f\!f'\!f'\!'}\ \tr[U_{\!f\!f'}
%-U_{\!f'\!f}
U^{-1}_{f'\!'}\{U_{f'\!'},V_t\}] \sim 0
\ee 
and then defining the corresponding quantum operator by replacing the Poisson bracket with the commutator.  Here the sum is over all ordered triples of distinct $f$'s and $\epsilon^{ff'f'\!'}$
is the parity of the ordered triple. 
The second possibility is to write directly the quantum operator  which corresponds to the regularization of the Hamiltonian constraint used earlier in loop quantum gravity \cite{smolincr}
\be\label{ham}
\tilde C_t = V_t C_t = \sum_{ff'\in t} \tr[U_{\!f\!f'}E_{f'}E_{f}
%-U_{\!f'\!f}E_fE_{f'}
] \sim 0. 
\ee
This form is more handy in the present context.  Multiplying (\ref{ham3}) by $V_t$, we can rewrite the full quantum constraint equations in the form
\be\label{hamfin}
S_t \psi=  \left(\frac{\kappa}{2}\  p_{\phi_t}^2 +  \tilde C_t \right)\psi = 0,  
\ee
This set of $n$ Hamiltonian constraints can be combined into a single one, introducing a ``Lapse" $N=\{N_t\}$ and writing 
\be\label{lapse}
S(N) \psi \equiv \sum_t N_t S_t\  \psi=0, \hspace{3em} \forall N.
\ee

This concludes our definition of a family of finite-dimensional inhomogeneous quantum cosmologies.  We have one of these for each triangulation $\Delta_n$ of a three-sphere.  The hypothesis that we put forward is that they give an approximate description of the quantum behavior of our inhomogeneous universe, increasingly accurate with $n$.  This hypothesis can be seen as following naturally from the cosmological principle. 

\section{``Dipole" cosmology}

Consider the simple case obtained by taking $n=2$ and the natural triangulation of the a three-sphere obtained by gluing two tetrahedra by all their faces.\footnote{%
An interesting possibility, which we leave to the reader, is also to consider the $n=1$ case defined by the graph
$\Delta_1^*\  = \begin{picture}(25,10)
\put(6,2) {\circle{10}}
\put(17,2) {\circle{10}}
\put(11.5,2) {\circle*{3}} 
\end{picture} $; namely a single tetrahedron with two couples of faces identified.}
  $\Delta_2^*$ is then the graph formed by two nodes joined by four links 
\begin{center}
\begin{picture}(20,40)
\put(-36,17) {$\Delta_2^*\  =$}
% cerchi 
\put(20,20) {\circle{30}}
\put(04,20) {\circle*{4}} 
\put(36,20) {\circle*{4}}  
% linee 
\qbezier(4,20)(20,33)(36,20)
\qbezier(4,20)(20,7)(36,20)
\end{picture} 
\end{center} 
The gravitational variables are $(U_f, E_f), f=1,2,3,4$.  We have two Hamiltonian constraints, one per each node, which we call $C_1$ and $C_2$.  Define 
\be 
\tilde C = V_1 C_1 + V_2 C_2 \
\ee
and rewrite the constraints in the equivalent form
\ba
S &=& \tilde C+\frac\kappa{2}(p_{\phi_1}^2+p_{\phi_2}^2)\sim 0,
\\
D &=& (V_1 C_1 - V_2 C_2) \ + \frac\kappa{2}(p_{\phi_1}^2-p_{\phi_2}^2)\sim 0. 
\ea
But using (\ref{Umeno1}) and (\ref{Emeno1}), it is easy to see that 
\be 
V_1 C_1=V_2 C_2. 
\ee
Therefore $D$ reads simply 
\be
D =  p_{\phi_1}^2-p_{\phi_2}^2  \sim 0. 
\ee
which shows immediately that the Poisson bracket algebra between the
two hamiltonian constraints closes.

The gravitational Hilbert space is $L_2[SU(2)^4]$ and a basis of spin network states that solve the gauge constraint is given by the states $|j_f, \iota_t\rangle=|j_1,j_2,j_3,j_4,\  \iota_1, \iota_2\rangle$.  The action of one gravitational Hamiltonian constraint on a state gives 
\be\label{ham4}
\tilde C |j_f, \iota_t\rangle = \sum_{ff'} C_{ff'}|j_f, \iota_t\rangle 
\ee
where, each term of the sum comes from one of the terms in the sum in $f$ and $f'$ in (\ref{ham}). 
More explicitly, we have 
\be
C_{12}|j_1,j_2,j_3,j_4, \iota_1, \iota_2\rangle 
=\sum_{\epsilon,\delta=\pm 1}
C^{\epsilon\delta \iota'_1\iota'_2 }_{j_f\iota_1\iota_2} \ |j_1+\frac\epsilon{2},j_2+\frac\delta{2},j_3,j_4, \iota'_1, \iota'_2\rangle, 
\ee
because the operator $U_{12}=U_1U_2^{-1}$ in (\ref{ham}) multiplies the terms $\Pi^{j_1}(U_1)$ and $\Pi^{j_2}(U_2)$ and
\be
U\Pi^{j}(U)= \Pi^{1/2}(U)\Pi^{j}(U)= c_+\Pi^{j+1/2}(U)+c_-\Pi^{j-1/2}(U).
\ee
The matrix elements $C^{\epsilon\delta \iota'_1\iota'_2 }_{j_f\iota_1\iota_2} $ can be computed with a straightforward exercise in recoupling theory from (\ref{ham}), and with some more algebra, from (\ref{hamth}).  In a different notation, in terms of the wave function components, we can write
\be
\tilde C\, \psi(j_f, \iota_t)\  
=\sum_{\epsilon_j=0,\pm 1}
{C}^{\epsilon_f \iota_t'}_{j_f \iota_t}\  \psi\!\left(j_f+\frac{\epsilon_f}{2}, \iota'_t\right), 
\ee
where $C^{\epsilon_j \iota_t'}_{j_f \iota_t}$ vanishes unless $\epsilon_f=0$ for two and only two of the four $j$'s.  
The scalar field variables are $\phi_1, \phi_2$. Taking these into account leads to the wave functions $\psi(j_f, \iota_n, \phi_n)$, and (\ref{hamfin}) gives the dynamical equations 
\ba\label{h1}
\frac\kappa{2} \left( \frac{\partial^2}{\partial \phi_1^2}
 + \frac{\partial^2}{\partial \phi_2^2}\right)\psi(j_f, \iota_t, \phi_t)&=&
 \sum_{\epsilon_f=0,\pm 1}
C^{\epsilon_f \iota_t'}_{j_f \iota_t}\  \psi\!\left(j_f+\frac{\epsilon_j}{2}, \iota'_t,\phi_t\right),\\
\label{h2}
\frac{\partial^2}{\partial \phi_1^2}\psi(j_f, \iota_t, \phi_t)&=&
\frac{\partial^2}{\partial \phi_2^2}\psi(j_f, \iota_t, \phi_t).
\ea
The coefficients ${ {C}}$ can be computed explicitly from recoupling theory.  They vanish unless two $\epsilon_f$'s are zero.  Equations (\ref{h1},\ref{h2}), defined on Hilbert space $H_2=L_2[SU(2)^4/SU(2)^2]\otimes L_2[R^2]$ define a quantum cosmological model which is just one step out of homogeneity. 

\section{Born-Oppenheimer approximation and LQC}

We now ask if and how LQC is contained in the model defined above.  
The state space $H_2$ contains a subspace that could be identified as a homogeneous universe.  This is the subspace $H^{hom}\subset H_2$ spanned by the states $|j,j,j,j, \iota_j,\iota_j,\phi,\phi\rangle$ where $\iota_j$ is the eigenstate of the volume that better approximates the volume of a classical tetrahedron whose triangles have area $j$.  However, the dynamical equations (\ref{h1},\ref{h2}) do not preserve this subspace. This is physically correct, because the inhomogeneous degrees of freedom cannot remain sharply vanishing in quantum mechanics, due to Heisenberg uncertainty.  Therefore it would be wrong to search for states that reproduce LQG \emph{exactly}, within this model.   In which sense then can a quantum homogeneous cosmology make sense? 

On the basis of the above discussion on the cosmological principle, the answer should be clear.   The cosmological principle is the hypothesis that in the theory there is a regime where the inhomogeneous degrees of freedom do not affect too much the dynamics of the homogeneous degrees of freedom, and that the state of the universe happens to be within such a regime. In other words, the homogeneous degrees of freedom can be treated as ``heavy" degrees of freedom, in the sense of the Born-Oppenheimer approximation, and the inhomogeneous one can be treated as ``light" ones. Let us therefore separate explicitly the two sets of degrees of freedom. 
This can be done as follows.

First, change variables from the group variables $U_f\in SU(2)$ to algebra variables $A_f\in su(2)$,  defined by $\exp A_f=U_f$.  Following  what is done in loop quantum cosmology \cite{ashtekarprimo},  let us fix a fiducial $su(2)$ element $\omega_{\!f}\in su(2)$ for each face $f$. (This can be interpreted as the logarithm of the holonomy of the fiducial connection along the link dual to $f$.)   We choose for simplicity a fiducial connection normalized as $|\omega_f|=1$, and
such that the four vectors $\omega_{\!f}$ are normal to the faces of a regular tetrahedron centered at the origin of $su(2)\sim R^3$. 
Using this, we can decompose our variables as
\ba
A_f&=&c\ \omega_{\!f}+a_f, \\
E_f&=&p\ \omega_{\!f}+h_f.  \label{din}
\ea 
We need two conditions in order to fix this decomposition uniquely. First, we require that $p$ is determined by the total volume
\be\label{Vp}
V= p^{\frac32}.
\ee 
Second, we require that $c$ is its conjugate variable, that is 
\ba\label{fixa}
\{c,p\} &=& \frac{8\pi G}3 = 1. 
\ea 
The variable $c$ can then be identified with the corresponding variable using in quantum cosmology.  We also define $\Delta V=V_2-V_1$, so that $V_{1,2}=\frac12(V\pm\Delta V)$. 

Inserting the decomposition described above in the quantum Hamiltonian constraint (\ref{hamth})  gives 
\be\label{hdec}
C_t =\frac{1}{2}\sum_{{f\!f'\!f'\!'\in t}} \tr\left[%\left(
e^{c\omega_{\!f}+a_{f}}e^{-c\omega_{\!f'}-a_{f'}}
%-e^{-c\omega_{\!f}-a_{f}}e^{c\omega_{\!f'}-a_{f'}}\right)
e^{-c\omega_{\!f'\!'}-a_{f'\!'}} [e^{c\omega_{\!f'\!'}+a_{f'\!'}},V\pm\Delta V]\right].
\ee 
Let us now decompose this constraint into two parts, the first of which depends only on the homogeneous variable $c$.  This can be done keeping only the first term of the expansion of the exponentials in $a_f$ and $a_{f'}$, and only the $V$ term in the volume term. That is, we write 
\be\label{hdec2}
C_t =\frac{1}{2}C^{hom}+C_t^{in}
\ee 
where
\be\label{hhom}
 C^{hom} =\sum_{{f\!f'\!f'\!'\in t}} \tr\left[%\left(
 e^{c\omega_{\!f}}e^{-c\omega_{\!f'}}
%-e^{-c\omega_{\!f}}e^{c\omega_{\!f'}}\right)
e^{-c\omega_{\!f'\!'}}
[e^{c\omega_{\!f'\!'}},V]\right]  \  \equiv   \ \  \frac1{V} \tilde C^{hom}.
\ee 
The interpretation of this spilt is transparent: $C^{hom}$ gives the gravitational energy in the homogeneous degree of freedom, while  $C_t^{in}$ gives the sum of the energy in the 
inhomogeneous degrees of freedom and the interaction energy between the two sets of degrees of freedom.  Finally, we write the homogeneous variable $\phi=\phi_1+\phi_2$ and $\phi_-=\phi_1-\phi_2$.

Following Born and Oppenheimer, let us now make the hypothesis that the state can be rewritten in the form
\be\label{bh}
\psi(U_f,\phi_t)=\psi_{hom}(c,\phi)\ \psi_{inh}(c,\phi;a_f,\phi_-), 
\ee
where the variation of $\psi_{inh}$ with respect to $c$ and $\phi$ can be neglected at first order. Here $\psi_{hom}$  represents the quantum state of the homogeneous cosmological variables, while $\psi_{inh}$ represents the quantum state of the inhomogeneous fluctuations over the homogeneous background $(c,\phi)$. 
Inserting the Born-Oppenheimer ansatz (\ref{bh}) into (\ref{lapse}), and taking $N_1=N_2$, we have the equation
\ba\label{din22}
\frac{\kappa}2\ \psi_{inh}
\frac{\partial^2}{\partial \phi^2}\psi_{hom}
+\frac{\kappa}2\  \psi_{hom}
\frac{\partial^2}{\partial \phi_-^2}\psi_{in}
-\psi_{in} \tilde C^{hom}\psi_{hom}-  \tilde C^{in}\psi_{hom}\psi_{in}=0.
\ea
Dividing by $\psi_{on}\psi_{inh}$ this gives 
\ba\label{din33}
\frac{\frac{\kappa}2\ \frac{\partial^2}{\partial \phi^2}\psi_{hom}}{\psi_{hom}} -
\frac{  \tilde C^{hom}\psi_{hom}}{\psi_{hom}}
&=&
-\frac{\frac{\kappa}2\ \frac{\partial^2}{\partial \phi_-^2}\psi_{inh}}{\psi_{inh}} + 
\frac{  \tilde C^{inh}\psi_{hom}\psi_{inh}}{\psi_{hom}\psi_{inh}}.
\ea
Since the left hand side of this equation does not depend on the inhomogeneous variables,
there must be a function $\rho(c,\phi)$ such that 
\ba\label{din44}
&& \frac{\kappa}2\ \frac{\partial^2}{\partial \phi^2}\psi_{hom} -
  \tilde C^{hom}\psi_{hom}- \rho{\psi_{hom}}=0, \\ 
&&\ \ \frac{\kappa}2\ \frac{\partial^2}{\partial \phi_-^2}\psi_{inh} + 
\frac{  \tilde C^{inh}\psi_{hom}\psi_{inh}}{\psi_{hom}}=\rho{\psi_{inh}}.
\ea
The second equation is the Schr\"odinger equation for the inhomogeneous modes in the background homogeneous cosmology $(c,\phi)$, where $\rho(c,\phi)$ plays the role of energy eigenvalue. The first equation is the quantum Friedmann equation for the homogeneous degrees of freedom $(c,\phi)$, corrected by the energy density $\rho(c,\phi)$ of the inhomogeneous modes \cite{B}.  At the order zero of the approximation, where we disregard entirely the effect of the inhomogeneous modes on the homogeneous modes, we obtain 
\be\label{questa}
\frac{\kappa}2\ \frac{\partial^2}{\partial \phi^2}\psi_{hom} 
=   \tilde C^{hom}\psi_{hom}.
\ee
Let us now analyze the action of the operator $C^{hom}$, defined in (\ref{hhom}). Notice that $c$ multiplies the generator of a $U(1)$ subgroup of $SU(2)^4$. Therefore it is a periodic variable $c\in[0,4\pi]$.  We can therefore expand 
the states $\psi_{hom}(c,\phi)$ in Fourier sum
\ba  
\psi_{hom}(c,\phi) &=& \sum_v \ \psi(v,\phi) \ e^{i\mu c/2}. 
\ea 
The basis of states $\langle c\,|\mu\rangle=e^{i\mu c/2}$ satisfies 
\ba  
p^{\frac32}|\mu\rangle&=& (\mu/2)^{\frac32}|\mu\rangle\\ 
-4\sin^2(c/2)|\mu\rangle&=&|\mu+2\rangle-2|\mu\rangle+|\mu-2\rangle
\label{s2}
\ea 
which we shall use below. 
% Here $k=\left(\frac{8\pi G}6\right)^{\frac32}$.
The homogeneous Hamiltonian constraint (\ref{hhom}) can be rewritten as 
\ba\label{hhom5}
C^{hom} &=&\sum_{\!f\!f'\!f'\!'} \tr\left[
(\cos{\frac c2}1\!\!1+2\sin{\frac c2}\omega_{\!f})(\cos{\frac c2}1\!\!1-2\sin{\frac c2}\omega_{\!f'})
%\right.\nonumber \\&&    \hspace{5em} \ \ \left. -(\cos{\frac c2}1\!\!1-2\sin{\frac c2}\omega_{\!f})(\cos{\frac c2}1\!\!1+2\sin{\frac c2}\omega_{\!f'}))\ \ 
e^{-c\omega_{\!f'\!'}}[e^{c\omega_{\!f'\!'}}, V]\right]\nonumber\\
%&=&\frac{1}{12}\sum_{\!f\!f'\!f'\!'} \tr\left[\left(2\sin{\frac c2}\cos{\frac c2}\; (\omega_{\!f}-\omega_{\!f'})-4\sin^2\!{\frac c2}\; [\omega_{\!f},\omega_{\!f'}]\right) e^{-c\omega_{\!f'\!'}}[e^{c\omega_{\!f'\!'}}, V]\right]. 
&=&\sum_{\!f\!f'\!f'\!'} \tr\left[\left(\cos^2{\frac c2}1\!\!1+2\sin{\frac c2}\cos{\frac c2}\; (\omega_{\!f}-\omega_{\!f'})+% giusto pi? perch c'era un meno?
4\sin^2\!{\frac c2}\; \omega_{\!f}\,\omega_{\!f'}\right) e^{-c\omega_{\!f'\!'}}[e^{c\omega_{\!f'\!'}}, V]\right]. 
\ea 
Consider the action of the last factor on the state $|\mu\rangle$
\ba\label{hhom55}
e^{-c\omega_{\!f'\!'}}[e^{c\omega_{\!f'\!'}}, V] e^{i\mu c/2}&=&
 p^{\frac32} e^{i\mu c/2}-e^{-c\omega_{\!f'\!'}}
 p^{\frac32}
 e^{c\omega_{\!f'\!'}}
e^{i\mu c/2}\nonumber\\
&=&
\left(-i\frac{8\pi G}3\frac{\partial}{\partial c}\right)^{\frac32} e^{i\mu c/2}
-e^{-c\omega_{\!f'\!'}} \left(-i\frac{8\pi G\gamma}3\frac{\partial}{\partial c}\right)^{\frac32} 
e^{ic(\mu/2-i\omega_{\!f'\!'})}\nonumber\\
&=&
 (\mu/2)^{\frac32} e^{i\mu c/2}
-e^{c\omega_{\!f'\!'}}k \left((\mu/2) 1\!\!1-i\omega_{\!f'\!'}\right)^{\frac32} 
e^{ic(\mu/2-i\omega_{\!f'\!'})}\nonumber\\
&=&
\left((\mu/2)^{\frac32}1\!\!1-\left((\mu/2)1\!\!1-i\omega_{\!f'\!'}\right)^{\frac32} 
\right) e^{i\mu c/2}.
\ea 
Now observe that we can write 
\be\label{equazione}
\left((\mu/2)1\!\!1-i\omega_{\!f'\!'}\right)^{\frac32} =
\alpha(\mu)1\!\!1+\beta(\mu)\omega_{\!f'\!'}
\ee 
where the coefficients $\alpha(\mu)$ and $\beta(\mu)$ can be easily computed squaring this equation.
%, using $\omega_{\!f'\!'}^2=-\frac14 1\!\!1$ and solving the resulting system, which gives 
%\ba\beta(\mu)=-\sqrt{-2\mu(\mu^2+3)+2(\mu^2-1)^{\frac32}} & \ \ \ \ \ \ & \alpha(\mu)=...\ .\ea
%(The two signs in the solution of the fourth degree equation can be determined directly from (\ref{equazione}) taking the large $\mu$ limit. )  
We write $\tilde{\alpha}(\mu) = (\mu/2)^{\frac32} - \alpha(\mu)$. Bringing everything together%
\footnote{We are using $\tr[\omega_f]=0  \, , \,  \tr [\,\omega_{f}\omega_{f'}] = -\f12 e^a_{f}e^b_{f'}q_{ab} = -\f12 \cos\theta_{ff'} $ and the symmetry of the homogenous case for which $ \cos\theta_{ff'} = \cos\theta_{ff''} = \f13 $ and 
$\tr \left[ \omega_{f} \,\omega_{f'} \, \omega_{f''} \right] = \, 3^{\f14} \, 2^{\f12} \,  %\frac {2\sqrt{2}} {3^{\f74}}
$ .}
the only term that survives is 
\ba\label{hhom54}
C^{hom} e^{i \mu c/2}&=& 
 \left[ \tilde{\alpha}(\mu) + \left(  \f74\tilde{\alpha}(\mu) + \, 3^{\f14} \, 2^{\f{5}{2}} \,  \beta(\mu) \right)\sin^2 \f c2\right]
%\tr\left[[\omega_{\!f},\omega_{\!f'}]\omega_{\!f'\!'}\right]
e^{i\mu c/2} \nn
% il seguente passaggio pu˜ essere omesso nell'articolo:
% &=& \f{1}{48} \left[ \tilde{\alpha}(\mu) + \left(  \f74\tilde{\alpha}(\mu) + \, 3^{\f14} \, 2^{\f{11}{2}}  \,  \beta(\mu) \right)\left( - e^{+i c} - e^{-i  c} +2 \right)\right] e^{i\mu c/2} \nn
% &=& \f{1}{48} \left[ - \left(  \f74\tilde{\alpha}(\mu) +  \, 3^{\f14} \, 2^{\f{11}{2}} \,  \beta(\mu) \right) \left(  e^{+i c} + e^{-i  c} \right) + \left(  \f{11}{2}\tilde{\alpha}(\mu) + 8  \, 3^{\f14} \, 2^{\f72} \,  \beta(\mu) \right) \right] e^{i\mu c/2} \nn
&=&  \left[{D}^+(\mu) \, e^{+i c} + {D}^-(\mu) \, e^{-i  c} + {D}^0 (\mu)\, %\mathbb{I}
 \right] e^{i\mu c/2} 
\ea 
where ${D}^+(\mu) \, = \,  {D}^-(\mu) \, = \,    -  \frac{7}{16}\,\tilde{\alpha}(\mu) -   \sqrt{2\sqrt{3}} \,  \beta(\mu)
\ \ \mbox{and}  \ \
{D}^0(\mu) \, = \,  \f{11}{8}\tilde{\alpha}(\mu) + \, 2\sqrt{2\sqrt{3}} \,  \beta(\mu)  $. \\ Using (\ref{s2}), this gives
\ba\label{hhom6}
C^{hom} |\mu\rangle\ &=&\ {D}^+(\mu)\ |\mu+2\rangle \ + \  {D}^0(\mu)\ |\mu\rangle \ Ê+
 \ {D}^-(\mu)\ |\mu-2\rangle . 
\ea 
%It follows that \ba\label{hhom6} \tilde C^{hom}_t |\mu\rangle &=&V C^{hom}_t |\mu\rangle \nonumber\\ &=& p^{\frac32} C\beta(\mu)\left(|\mu+2\rangle-2|\mu\rangle+|\mu-2\rangle\right) \nonumber\\ &=&  kC\beta(\mu)\left((\mu+2)^{\frac32}|\mu+2\rangle-2\mu^{\frac32}|\mu\rangle+(\mu-2)^{\frac32}|\mu-2\rangle\right). \ea 
The full equation (\ref{questa}) can be written as
\ba\label{din2}
 \ \frac{\partial^2}{\partial \phi^2}\Psi(\mu,\phi) &=& C^+(\mu)\ \Psi(\mu+2,\phi)+C^0(\mu)\ \Psi(\mu,\phi)+C^-(\mu)\ \Psi(\mu-2,\phi). 
\ea
where we have written $\Psi$ for $\psi_{hom}$ and $C^{\pm,0}(\mu)=\frac{\mu^{3/2}}{\kappa\sqrt{2}}\,D^{\pm,0}(\mu) $.
Equation (\ref{din2}) has the structure of the LQC dynamical equation.  Thus, LQC appears in the zero order Born-Oppenheimer approximation of a loop quantum gravity quantization of a finite number of degrees of freedom of the gravitational field, truncated according to the approximation dictated by the cosmological principle. 

\section{Conclusions and perspectives}

We leave a number of questions open.  In particular: (i) the extension to real Immirzi parameter $\gamma$; (ii) the inclusion of more realistic matter fields; (iii) understanding the relation between the $\psi(v,\phi)$ homogeneous states and the full $\psi(j_f,\iota_t,\phi_t)$ states in the spinnetwork basis; and (iv) the detailed comparison with the LQC quantization of the homogeneous universe, in particular in relation to the $\bar\mu$ quantization scheme \cite{ashtekar, Corichi:2008zb}.  The relation with more conventional cosmological perturbation expansion (see \cite{Dittrich:2007jx}) need also to be investigated. 

We observe that the discussion on the cosmological principle given in Section 2 and the observation on the use of coarse triangulations in Section 3 bear also on the discussion on the viability of the recent computations of background independent $n$-point functions \cite{propagator}. The conclusion of the discussion was the possibility of representing a large universe in terms of a coarse triangulation, and therefore, in the quantum theory, in terms of states based on graph with a small numbers of nodes. In cosmology, as in the calculation of the $n$-point functions, we observe that general relativity admits an expansion in which a region of length scale $L$ can be approximated by neglecting wavelengths much smaller than $L$, and described by spin networks with a small number of nodes in the quantum theory.  In other words, the intuition that dynamics of nearly flat space can only be described using states with a high number of nodes is misleading. If this was the case, cosmology itself would be ill conceived. 

The idea of describing cosmological evolution using Regge calculus has been explored in the past.  See for instance \cite{barrett}. Here we have adapted this idea to quantum cosmology, where it turns out to be particularly suitable for a loop quantization.  The construction in \cite{barrett} indicates that it is possible to have a 4d triangulation sliced by 3d triangulations equal to one another, and therefore suggests the possibility of writing a spinfoam version of the models introduced here.   For instance, consider the kinematics of the $\Delta_2$ model and let $A(j_{ab}, \iota_a), a=1,...,5$ be the vertex amplitude of a spinfoam model, as for instance that introduced in \cite{EPR}. We can interpolate between two $\Delta_2$ with the triangulation $\Delta_5$ defined by the boundary of a four-simplex.  In fact, collapsing four of the five tetrahedra of $\Delta_5$ into a single one gives precisely $\Delta_2$. This collapse is a 4-1 Pachner move, which can be realized interpolating a four-simplex.  Therefore we can have a transition $\Delta_2\to\Delta_2$ via an intermediate $\Delta_5$.  See Figure 1.

\begin{figure}[h]
\centering
\includegraphics[scale=0.2]{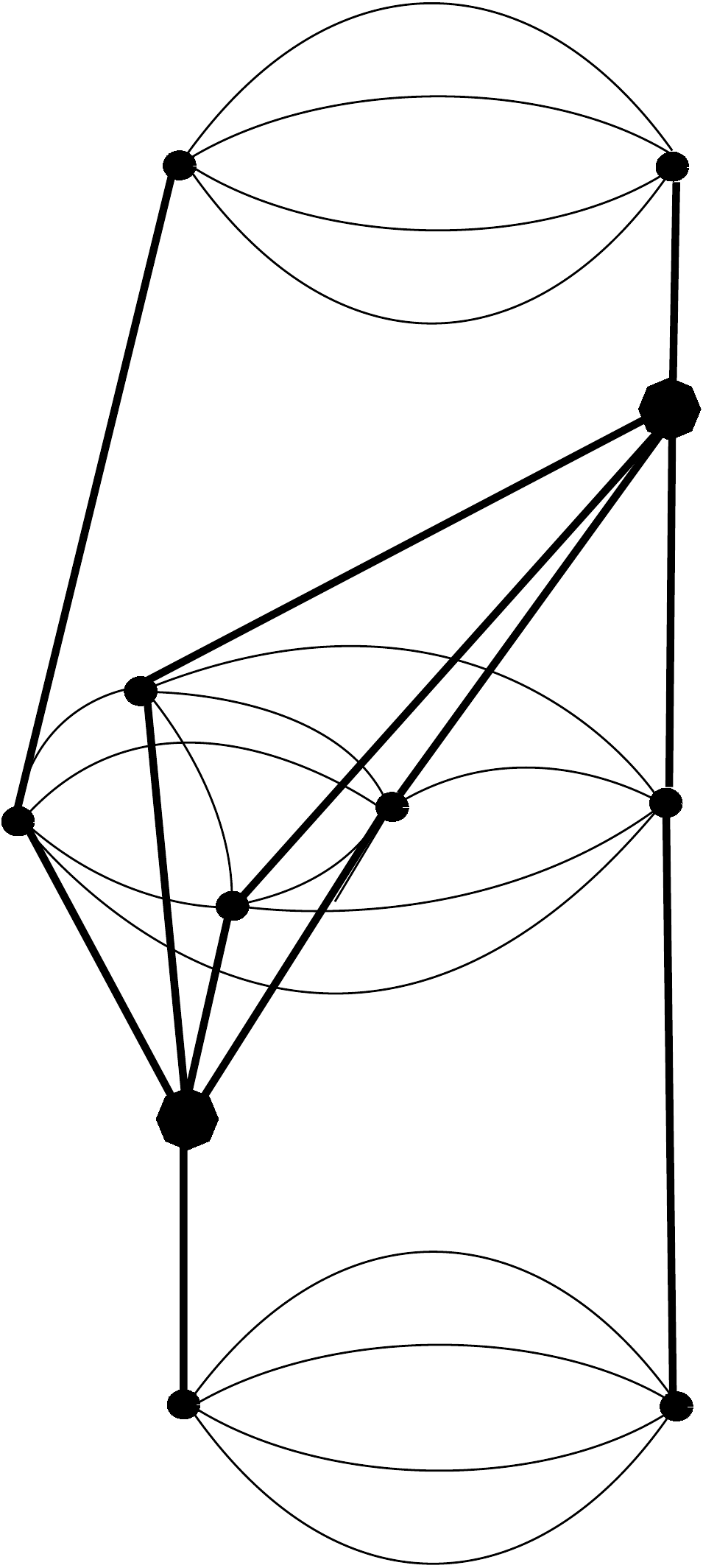}
\label{propagator}
\caption{Spinfoam evolution of the dipole cosmology.}
\end{figure}

Accordingly, we can write a $\Delta_2\to\Delta_2$ transition amplitude as 
the $\Delta_2\to\Delta_5\to\Delta_2$ amplitude defined by the transition amplitude $A(j_{ab}, \iota_a)$ for each move. For instance
\be
A(j_f,\iota_t;j'_f,\iota'_t)=\delta_{j_1,j'_1}\sum_{\iota_c,j_c} A(j_t,j_c,j'_t,\iota_c, \iota_1,\iota'_1)A(j_t,j_c,j'_t,j_{ab}, \iota_c, \iota_2,\iota'_2).  
\ee
where $c=1,2,3$. Repeating this step four times, over different points of the triangulation generates spacetime with the $S_3\times [0,1]$ topology \cite{barrett}.   Is this dynamics related to the canonical one defined here?  

In summary, the models presented in Section 3 open a systematic way for describing the inhomogeneous degrees of freedom in quantum cosmology.  In particular, they open the possibility of checking whether the bounce scenario that is characteristic of the homogeneous theory survives in a inhomogeneous context.  The simplest possibility is to analyze the cosmological evolution in the simple $\Delta_2$ model described in Section 4, defined by the two equations (\ref{h1}) and (\ref{h2}). Do these equations govern semiclassical wave packets undergoing the cosmological bounce?  If the answer to the above question is positive, the solution would also provide a concrete description of the fluctuating geometry at the bounce. The state $\psi(j_f,i_t,\phi_t=\phi_{bounce})$ would give such a description explicitly.  It is tempting to begin speculate on the possible cosmological role of the fluctuations of the inhomogeneous degrees of freedom at (or near) the bounce.  Could they play a role in structure formation? For inflation?

The derivation of the structure of the LQC dynamical equation presented in Section 5 sheds light on the relation between LQC and full loop quantum gravity.  In particular, we have argued that the first should be searched as a Born-Oppenheimer approximation, as suggested by the hypothesis that goes under the name of cosmological principle.

Finally, we point out the existence of the term $\rho(c,\phi)$ in the quantum Friedmann equation (\ref{din}). Its physical interpretation is clear: it represents the back reaction of the quantum fluctuating inhomogeneous degrees of freedom on the dynamics of the scale factor. Again, it is tempting to begin to speculate on the possible cosmological role of this energy density.  Does it play a role in structure formation? For inflation? In relation to the cosmological constant? 

\vspace{.5cm}
\centerline{------}
\vspace{.5cm}

We thank the audience of the ILQG seminar, where this work was first presented, for the useful questions and an enlightening discussion.  We thank Bianca Dittrich, John Barrett and Simone Speziale for useful exchanges. 

\newpage

\appendix
\section{A simpler Hamiltonian constraint}

Instead of using the form (\ref{hamth}) of the Hamiltonian constraint, we can also use the simpler
densitized form (\ref{ham}) which corresponds to the regularization of the Hamiltonian constraint used earlier in loop quantum gravity in \cite{smolincr}. Here we show that this regularization gives a simpler dynamical equation, but with the same structure as in LQC. 

In the quantum theory, the decomposition (\ref{din}) becomes 
\be\label{Ldec}
L_f= \omega_{\!f} \frac{\partial}{\partial c} +   \tilde L_f . 
\ee 
where $\tilde L_f c =0$. 
Inserting this decomposition into the Hamiltonian constraint (\ref{ham}) gives 
\be\label{hdec3}
\tilde{C_t} =\sum_{ff'\in t} \tr\left[%\left(
e^{c\omega_{\!f}+a_{f}}e^{-c\omega_{\!f'}-a_{f'}}
%-e^{-c\omega_{\!f}-a_{f}}e^{c\omega_{\!f'}-a_{f'}}\right)
\left( \omega_{\!f'} \frac{\partial}{\partial c} +   \tilde L_f \right)
\left(\omega_{\!f} \frac{\partial}{\partial c} +   \tilde L_{f'}\right)\right].
\ee 
As before, we us decompose this constraint in two parts, the first of which depends only on the homogeneous variable $c$.  This can be done keeping only the first term of the expansion of the exponentials in $a_f$ and $a_{f'}$, and only the term quadratic in $\frac{\partial}{\partial c}$ in the derivative part. That is
\be\label{ahhom}
\tilde{C_t}^{hom} =\sum_{ff'\in t} \tr[e^{c\omega_{\!f}}e^{-c\omega_{\!f'}}
%-e^{-c\omega_{\!f}}e^{c\omega_{\!f'}}) 
\omega_{\!f'} \omega_{\!f}] \frac{\partial}{\partial c}
 \frac{\partial}{\partial c}\  \  \equiv     \ \  \frac12 C^{hom} .
\ee 
This can be rewritten as 
\ba\label{ahhom555}
\tilde{C}^{hom}&=&\sum_{\!f\!f'} \tr[
(\cos{\frac c2}1\!\!1+2\sin{\frac c2}\omega_{\!f})(\cos{\frac c2}1\!\!1-2\sin{\frac c2}\omega_{\!f'})
\omega_{\!f'} \omega_{\!f}] \frac{\partial^2}{\partial c^2}
\nn
%&=&\sum_{\!f\!f'} \tr\left[ \left(\cos^2{\frac c2}1\!\!1+2\sin{\frac c2}\cos{\frac c2}\; (\omega_{\!f}-\omega_{\!f'})+ 4\sin^2\!{\frac c2}\; \omega_{\!f}\,\omega_{\!f'}\right)\omega_{f'}\omega_{\!f}\right] \frac{\partial^2}{\partial c^2} \nn
&=&\f14 (-e^{ic}+1-e^{-ic}) \frac{\partial^2}{\partial c^2}. 
\ea 
The action of this operator on the states $\psi_{hom}(\mu,\phi)$ is therefore easily computed
\be\label{ahhom3}
\tilde{C}^{hom}\psi_{hom}(v,\phi)= \f14 [
-\mu^2 \psi_{hom}(\mu+2,\phi) 
+\mu^2 \psi_{hom}(\mu,\phi) 
-\mu^2 \psi_{hom}(\mu-2,\phi)].
\ee 
Bringing everything together, the full equation (\ref{questa}) reads 
\be
C^+(\mu)\ \psi_{hom}(\mu+2,\phi)+C^0(\mu)\ \psi_{hom}(\mu,\phi)+C^-(\mu)\ \psi_{hom}(\mu-2,\phi)+ \frac{\partial^2}{\partial \phi^2}\psi_{hom}(\mu,\phi) =0. 
\ee
where the coefficient take the simple form $C^\pm(\mu)= - C^0(\mu)=\frac{\mu^2}{4\kappa} $.  We do not know if this simple equation gives the same phenomenology as the one used in LQC. 
\vspace{.5cm}
%\newpage


\begin{thebibliography}{99}

\bibitem{ashtekar}  

A.\,Ashtekar, M.\,Bojowald, J.\,Lewandowski ``Mathematical structure of loop quantum
cosmology", Adv. Theor. Math. Phys. 7 (2003) 233-268.
    
A.\,Ashtekar, T.\,Pawlowski, P.\,Singh, ``Quantum Nature of the Big Bang'', arXiv:gr-qc/0602086v2.

A.\,Ashtekar, A.\,Corichi, P.\,Singh, ``Robustness of key features of loop quantum cosmology'', arXiv: 0710.3565.

A.\,Ashtekar, ``An Introduction to Loop Quantum Gravity Through Cosmology'', arXiv:gr-qc\ 0702030.

\bibitem{bojowald} 

M.\,Bojowald,  ``Isotropic loop quantum cosmology", Class.\,Quant.\,Grav. 19 (2002) 2717-2742.

M.\,Bojowald, ``Loop Quantum Cosmology'', {Liv.\,Rev.\,Rel.} {8} (2005) 11.

\bibitem{Kiefer:2008sw}
  C.\,Kiefer and B.\,Sandhoefer,
  ``Quantum Cosmology", 
  arXiv:0804.0672.
  

\bibitem{fotini}
T.\,Konopka, F.\,Markopoulou,  L.\,Smolin,
``Quantum graphity",  hep-th/0611197.

T.\,Konopka, F.\,Markopoulou, S.\,Severini,
  ``Quantum Graphity: a model of emergent locality", 
  arXiv:0801.0861.

\bibitem{lqg}
C.\,Rovelli, {Quantum Gravity}, (Cambridge University Press, Cambridge, 2004).

A.\,Ashtekar, J.\,Lewandowski, ``Background independent quantum gravity: a
status report", Class.\ Quantum Grav. 21 (2004) R53.

T.\,Thiemann, ``Modern canonical quantum general relativity,'' (Cambridge
  University Press, Cambridge, UK, 2007).

 \bibitem{lqg2}  
C.\,Rovelli, L.\,Smolin,
 ``Knot theory and quantum gravity'' 
{Phys.\,Rev.\,Lett} {61} (1988) 1155-1158. %\\
%%CITATION = PRLTA,61,1155;%%

C.\,Rovelli, L.\,Smolin,
``Loop space representation for quantum general relativity", 
{Nucl.\,Phys.} {B331}  (1990) 80-152.%\\
 %%CITATION = NUPHA,B331,80;%%

 A.\,Ashtekar, C.\,Rovelli, L.\,Smolin,
 ``Weaving a classical geometry with quantum threads",
{Phys.\,Rev.\ Lett}  {69}, 237 (1992).
  %%CITATION = PRLTA,69,237;%%

C.\,Rovelli, L.\,Smolin,
 ``Discreteness of Area and Volume in Quantum Gravity", 
{Nucl.\,Phys.} {B442} (1995)  593-619;
 %%CITATION = GR-QC 9411005;%%
{Nucl.\,Phys.} {B456}, 734 (1995).%\\

  C.\,Rovelli,
 ``Loop quantum gravity and black hole physics,''
  Helv.\ Phys.\ Acta {69}, 582 (1996).
  %%CITATION = HPACA,69,582;%%
  
 A.\,Perez, ``Spin foam models for quantum gravity,'' {Class.\,Quant.\,Grav.}
 {20} (2003) R43. 

\bibitem{thiemann} 
M\,Bojowald, H.\,A.\,Kastrup,  ``Quantum symmetry reduction for diffeomorphism invariant
theories of connections ", Class. Quantum Grav. 17 (2000) 3009-3043.

M\,Bojowald, H.\,A.\,Kastrup,  ``Symmetric states in quantum geometry", gr-qc/0101061. 

T.\,Koslowski, ``Reduction of a quantum theory", gr-qc/0612138.

J.\,Engle, ``Black Hole Entropy, Constraints, and Symmetry in Quantum Gravity", Ph.D.
Thesis, Penn State University, http:/\!/igpg.gravity.psu.edu/archives/thesis/2006/englethesis.pdf.

J.\,Engle,
``On the physical interpretation of states in loop quantum cosmology'',  gr-qc/0701132.

J.\,Engle, ``Relating loop quantum cosmology to loop quantum gravity: Symmetric
sectors and embeddings", gr-qc/0701132.

\bibitem{fleishhach}  J.\,Brunnemann, C.\,Fleischhack, ``On the Configuration Spaces of Homogeneous Loop Quantum Cosmology and Loop Quantum Gravity'', arXiv:0709.1621v1 

\bibitem{MartinBenito:2008ej}
  M.~Martin-Benito, L.~J.~Garay and G.~A.~Mena Marugan,
  ``Hybrid Quantum Gowdy Cosmology: Combining Loop and Fock Quantizations", 
  arXiv:0804.1098 [gr-qc].
  %%CITATION = ARXIV:0804.1098;%%
  
   \bibitem{Dittrich:2007jx}
  B.~Dittrich and J.~Tambornino,
  ``Gauge invariant perturbations around symmetry reduced sectors of general
  relativity: Applications to cosmology,''
  Class.\ Quant.\ Grav.\  {24} (2007) 4543
  arXiv:gr-qc/0702093.
  %%CITATION = CQGRD,24,4543;%%

\bibitem{Bojoino}
M.~Bojowald, ``Loop quantum cosmology and inhomogeneities", 
Gen.\ Rel.\ Grav. 38 (2006) 1771-1795.   arXiv:gr-qc/0609034.

\bibitem{Bojoino2}
M.~Bojowald, D.~Cartin, G.~Khanna,
``Lattice refining loop quantum cosmology, anisotropic models and stability",
 Phys.\ Rev.\  D 76 (2007) 064018,  
arXiv:0704.1137.

M.~Bojowald, H.~Hernandez, M.~Kagan, P.~Singh, A.~Skirzewski, 
``Formation and Evolution of Structure in Loop Cosmology", 
 Phys.\  Rev.\  Lett.\   98 (2007) 031301, 
arXiv:astro-ph/0611685.

M.~Bojowald, R.~Swiderski, 
``Spherically Symmetric Quantum Geometry: Hamiltonian Constraint", 
 Class.\  Quant.\  Grav. 23 (2006) 2129-2154, 
arXiv:gr-qc/0511108.  

M.~Bojowald, H.~H.~Hernandez, H.~A.~Morales-Tecotl,  
``Perturbative Degrees of Freedom in Loop Quantum Gravity: Anisotropies", 
 Class.\  Quant.\  Grav. 23 (2006) 3491-3516, 
arXiv:gr-qc/0511058.
 
M.~Bojowald, 
``Spherically Symmetric Quantum Geometry: States and Basic Operators", 
 Class.\  Quant.\  Grav. 21 (2004) 3733-3753, arXiv:gr-qc/0407017. 
 
\bibitem{Ellis} G.\ F.\ R.\,Ellis, ``Relativistic Cosmology" , http://www.pv.infn.it/~spacetimeinaction/ \\ speakers/transparencies/Ellis/pdf/Ellis1.pdf, slide 58, Conference ``Spacetime in Action", Pavia 2005.

\bibitem{thiemanncit} T.\,Thiemann, ``The Phoenix Project: Master Constraint Programme for Loop Quantum Gravity'', {Class.\,Quant.\,Grav.} {23} (2006) 2211-2248.

\bibitem{algebraic}
K.\,Giesel, T.\,Thiemann, ``Algebraic Quantum Gravity (AQG) I. Conceptual Setup'',
arXiv:gr-qc/ 0607099. 

K.\,Giesel, T.\,Thiemann, ``Algebraic Quantum Gravity (AQG) IV. Reduced Phase Space Quantisation of Loop Quantum Gravity'', arXiv:0711.0119v1 [gr-qc]
%controllare, forse era il III e non il IV

\bibitem{BO} M.\,Born, R.\,Oppenheimer, ``Zur Quantentheorie der Molekeln", Annalen der Physik, 17 (1927) 457.

\bibitem{BOP} J.\,Halliwell, S.\,Hawking, ``Origin of structure in the universe", Phys Rev D 31  (1985) 1777-91. 

A.\,Vilenkin, ``Interpretation of the wave-function of the universe", Phys Rev D39 (1989) 1116-1122. 

C.\,Kiefer, ``The semiclassical approximation to QG", in {\em Canonical Quantum Gravity: from classical to quantum}  J.\,Ehlers, H.\,Friedrich editors, pp. 170-212 (Springer, Berlin, 1994).

C.\,Kiefer, {\em Quantum Gravity}  (Oxford University Press, Oxford 2007).



\bibitem{barrett} 
J.W.\,Barrett, M.\,Galassi, W.A.\,Miller, R.D.\,Sorkin, P.A.\,Tuckey, R.M.\,Williams, ``A Parallelizable Implicit Evolution Scheme for Regge Calculus'', arXiv:gr-qc/9411008v1

\bibitem{propagator} 
L.\,Modesto, C.\,Rovelli, ``Particle scattering in loop quantum
gravity," Phys.\,Rev.\,Lett.\, 95 (2005), 191301. 
%%CITATION = GR-QC/0202017;%%.


 C.\,Rovelli, ``Graviton propagator from background--independent
quantum gravity", Phys.\,Rev. Lett.\,97 (2006), 151301.
 %%CITATION = GR-QC 0508124;%%

E. Bianchi, L. Modesto, C. Rovelli, S. Speziale, ``Graviton propagator
in loop quantum gravity," Class.\,Quant.\,Grav.\,23 (2006), 6989-7028.
   %%CITATION = GR-QC 0604044;%%

E.\,R.\,Livine, S.\,Speziale, ``Group integral techniques for the spinfoam
  graviton propagator,'' {JHEP} {11} (2006) 092. 
%%CITATION = GR-QC/0608131;%%.


S.\,Speziale, ``Towards the graviton from spinfoams: The 3d toy model,'' {
  JHEP} {05} (2006) 039. 
%%CITATION = GR-QC/0512102;%%.

E.\,R.\,Livine, S.\,Speziale,  J.\,L.\,Willis, ``Towards the graviton from
  spinfoams: Higher order corrections in the 3d toy model,'' {Phys.\,Rev.}
  {D75} (2007) 024038.
%%CITATION = GR-QC/0605123;%%.

E.\,Bianchi, L.\,Modesto, 
``The perturbative Regge-calculus regime of Loop Quantum Gravity", 
arXiv: 0709.2051.

E.\,Alesci, C.\,Rovelli, ``The complete LQG graviton propagator: I.\,Difficulties
with the Barrett-Crane vertex", Phys.\,Rev.\,D, to appear (2007), arXiv:07080883.
%%CITATION = ARXIV:0708.0883;%%.

\bibitem{einstein} A.\,Einstein, ``Kosmologische Betrachtungen zur allgemeinen RelativitŠtstheorie", Koniglich Preussische Akademie der Wissenschaften, 
Sitzungsberichte 1 (1917) 421.

\bibitem{B}
T.~Buchert,
  ``Dark Energy from Structure - A Status Report,''
  Gen.\ Rel.\ Grav.\ 40 (2008) 467. 
  
  T.~Buchert and M.~Carfora,
  ``On the curvature of the present-day Universe,''
  arXiv:0803.1401.
 
 F.\,Debbasch. ``What is a mean gravitational field?" Eur. Phys. J. B, 37 (2004) 257Ð270.  


\bibitem{smolincr}
C.\,Rovelli, L.\,Smolin, ``The physical 
Hamiltonian in nonperturbative quantum gravity", 
Phys.\,Rev.\ Lett.\,72  (1994) 446.


\bibitem{ashtekarprimo} A.\,Ashtekar,  ``Loop quantum cosmology of k=1 FRW models",  
gr-qc/0612104.

  \bibitem{EPR} J.\,Engle, R.\,Pereira, C.\,Rovelli, ``The loop-quantum-gravity vertex amplitude",  
Phys.\,Rev.\,Lett.\,99 (2007) 161301.
%%CITATION = ARXIV:0705.2388;%%.

J.\,Engle, R.\,Pereira, C.\,Rovelli, ``Flipped spinfoam vertex and loop gravity",  
arXiv:0708.1236. 
%%CITATION = ARXIV:0708.1236;%%.

E.\,Livine, S.\,Speziale, ``A New spinfoam vertex for quantum gravity", arXiv:0705.0674. 
%%CITATION = ARXIV:0708.1915;%%.

E.\,Livine, S.\,Speziale, ``Consistently Solving the Simplicity Constraints for Spinfoam Quantum Gravity",  arXiv:0708.1915. 

L.\,Freidel, K.\,Krasnov, ``A New Spin Foam Model for 4d Gravity", arXiv:0708.1595. 

S.\,Alexandrov, ``Spin foam model from canonical quantization,''
arXiv:0705.3892.

R.\,Pereira, ``Lorentzian LQG vertex amplitude", to appear.

J.\,Engle, E.\,Livine, R.\,Pereira, C.\,Rovelli, ``LQG vertex for general values of the 
Immirzi parameter", to appear.


\bibitem{tetraedro}
A\,Barbieri, ``Quantum tetrahedra and simplicial spin networks'',
 Nucl.\,Phys.\,B518 (1998) 714-728.
 
J.\,C.\,Baez, J.\,W.\,Barrett, ``The quantum tetrahedron in 3 and 4
  dimensions,'' {Adv.\,Theor.\,Math. Phys.} {3} (1999) 815--850.

A.\,Ashtekar, A.\,Corichi,  J.\,A.\,Zapata, ``Quantum theory of geometry.\,III:
  Non-commutativity of Riemannian structures,'' {Class.\,Quant.\,Grav.} {
  15} (1998) 2955--2972.
  

\bibitem{sc} C.\,Rovelli S.\,Speziale, ``A Semiclassical tetrahedron", Class.\,Quant.\,Grav.\,23 (2006) 5861-5870.
%%CITATION = GR-QC/0606074;%%.

\bibitem{Corichi:2008zb}
  A.\,Corichi and P.\,Singh,
  ``Is loop quantization in cosmology unique?", 
  arXiv:0805.0136.
  

\end{thebibliography}
\end{document}